\documentclass[onecollarge,natbib]{svjour2}
\bibpunct{[}{]}{;}{n}{}{,} 
\smartqed  
\usepackage{graphicx}
\usepackage{epstopdf}
\usepackage{braket}
\journalname{Archive of Applied Mechanics}
\begin{document}

\title{Transverse momentum distributions of electron in simulated QED model}



\author{Navdeep Kaur         \and
        Harleen Dahiya 
}


\institute{N. Kaur \and H. Dahiya
\\ Department of Physics, Dr. B. R. Ambedkar National Institute of Technology, Jalandhar - 144011, INDIA \\
              \email{nkmangat91@gmail.com}           
}

\date{Received: date / Accepted: date}

\maketitle

\begin{abstract}
In the present work, we have studied the transverse momentum distributions (TMDs) for the electron in simulated QED model. We have used the overlap representation of light-front wave functions (LFWFs) where the spin-1/2 relativistic composite system consists of spin-1/2 fermion and spin-1 vector boson. The results have been obtained for T-even TMDs in transverse momentum plane for fixed value of longitudinal momentum fraction $x$.
\end{abstract}

\section{Introduction}
\label{intro}
The electron is a structureless object but in the quantum theory, due to the Heisenberg uncertainty principle, an electron can fluctuate into a virtual electron-photon pair, carrying the same quantum number as an electron, i.e. $ e\rightarrow e\gamma \rightarrow e $. Further, these virtual photons can themselves break up into pairs of virtual electrons and positrons and, as a result, an isolated electron is treated as a composite system of virtual photons, electrons and positrons. Therefore, the bare electron effectively becomes a dressed electron with the photons, electrons, and positrons which can be interpreted as partonic structure of an electron
\cite{Bacchetta}.\\
In the scattering process, the partons interact with the probe and  reveal the internal structure of an electron. In analogy to the QCD, the partonic structure of an electron in QED can be described by different type of distribution functions like generalized parton distributions (GPDs), transverse momentum dependent distributions (TMDs), generalized transverse momentum dependent distributions (GTMDs) and Wigner distributions. In this paper, we have discussed the TMDs of the active electron in a dressed electron. TMDs describe the probability to find in a dressed electron a parton with longitudinal momentum $\textit{x}$ and transverse-momentum $\textit{k}_{T}$ with respect to the direction of the parent electron momentum. At leading twist, there are eight TMDs for active electron out of which six are T-even TMDs and remaining two are T-odd TMDs.
We have adopted the formalism of light-cone quantization to study parton distributions. In order to illustrate the relativistic features of composite system, we have used the QED model in which a spin-1/2 composite system consists of spin-1/2 fermion (internal electron) and spin-1 vector boson (photon) \cite{bacchetta, Meibner}. An overlap representation of TMDs in terms of the light-front wave functions (LFWFs) are used in this work. In simulated QED model of leptons, the behaviour of wave functions enhanced at the end points at $ x = 0, 1 $ as discussed in \cite{vary, asmita}. By differentiation of the QED LFWFs with respect to $ M^2 $, the simulated QED model has been obtained. This approach have been also used to obtained the GPDs of proton \cite{dahiya} and the QED form factors \cite{narinder}. We have evaluated the TMDs for an electron in this model.
\section{TMDs for an electron}
\label{sec:1}
In light-front quantizated QED, the anomalous magnetic moment of an electron is due to the one-fermion one-gauge boson lowest fock state component of a physical electron. So we have considered a dressed electron with momentum $\textit{P}$ and spin $\textit{S}$, composed of a bare electron of momentum $ k $ and a photon with momentum $ P - k $.
The internal electron TMDs are defined through the correlation function as follow \cite{Meibner, boer}:
\begin{eqnarray}
\label{eqn:5}
\Phi^{[\Gamma]}(x,\textit{\textbf{k}}_{\bot};S) = \frac{1}{2}\int\frac{dy^{-}d^{2}\textbf{\textit{y}}_{T}}{16\pi^3} e^{i\textit{k}.y} \langle P,S\vert \bar{\psi}(0)\psi(\textit{y})\vert P,S\rangle \big\vert_{{y}^{+}=0},
\end{eqnarray}
where $ \Gamma $ is the Dirac operator and $ \textit{P} = \big(\textit{P}^{+}, \textbf{0}, \frac{M^{2}}{ 2 \textit{P}^{+}}\big) $ defines the incoming dressed electron. At leading twist, the internal electron TMDs are obtained from the above correlator \cite{Bacchetta, Meibner} :
\begin{eqnarray}
\label{eqn:6}
\Phi(x,\textit{\textbf{k}}_{\bot};S) = \Phi^{[\gamma^{+}]}(x,\textit{\textbf{k}}_{\bot};S) &=& f_{1}^{e}(x,\textit{\textbf{k}}_{\bot}^{2})- \frac{\epsilon_{T}^{ij} \textit{k}_{\bot}^{i} S_{T}^{j}}{M} f_{1T}^{\perp e}(x,\textit{\textbf{k}}_{\bot}^{2}),\nonumber\\
\tilde{\Phi}(x,\textit{\textbf{k}}_{\bot};S) = \Phi^{[\gamma^{+} \gamma_{5}]}(x,\textit{\textbf{k}}_{\bot};S) &=& \Lambda g_{1L}^{e}(x,\textit{\textbf{k}}_{\bot}^{2}) + \frac{\textit{\textbf{k}}_{\bot}.\textbf{S}_{T}}{M} g_{1T}^{e}(x,\textit{\textbf{k}}_{\bot}^{2}),\nonumber\\
\Phi_{T}^{j}(x,\textit{\textbf{k}}_{\bot};S) = \Phi^{[i\sigma^{j+}\gamma_{5}]}(x,\textit{\textbf{k}}_{\bot};S) &=& -\frac{\epsilon_{T}^{ij}\textit{k}_{\bot}^{i}}{M} h^{\bot e}_{1}(x,\textit{\textbf{k}}_{\perp}^{2}) + \frac{\Lambda \textit{k}_{\bot}^{i}}{M} h_{1L}^{\bot e}(x,\textit{\textbf{k}}_{\bot}^{2})\nonumber\\
&+& S_{T}^{j} h_{1}^{e}(x,\textit{\textbf{k}}_{\bot}^{2}) + \frac{2\textit{k}_{\bot}^{j}\textit{\textbf{k}}_{\bot}.\textit{\textbf{S}}_{T}-S_{T}^{j}\textit{\textbf{k}}_{\bot}^{2}}{2M^{2}}h_{1T}^{\bot e}(x,\textit{\textbf{k}}_{\bot}^{2}).
\end{eqnarray}
In this paper, we have used $ \psi^{\Lambda}_{\lambda_{e},\lambda_{\gamma}} $ as general form of the LFWFs for an electron   with dressed electron helicity $ \Lambda $, internal electron helicity $ \lambda_{e} $ and photon polarization $ \lambda_{\gamma} $ which are constrained by ``spin sum rule", i.e. ${\Lambda} = \lambda_{\textit{e}} + \lambda_{\gamma} + \textit{L}_{\textit{z}} $. The $\textit{L}_{\textit{z}}$ is the projection of the relative orbital angular momentum between the internal electron and the photon. For the two-particle Fock state of an electron, $\textit{L}_{\textit{z}}$ has only 0 and  $\pm 1$ values which corresponds to S-wave and P-waves, respectively.\\
The overlap representation of the T-even TMDs in terms of LFWFs are \cite{Bacchetta},
\begin{eqnarray}
f_{1}^{e}(x,\textbf{k}_{\perp}^2) &=& \frac{1}{16\pi^3}[\mid\Psi^{\uparrow}_{+,+1}(x,\textbf{k}_{\perp})\mid^2 + \mid\Psi^{\uparrow}_{+,-1}(x,\textbf{k}_{\perp})\mid^2 + \mid\Psi^{\uparrow}_{-,+1}(x,\textbf{k}_{\perp})\mid^2],\nonumber\\
g_{1L}^{e}(x,\textbf{k}_{\perp}^2) &=& \frac{1}{16\pi^3}[\mid\Psi^{\uparrow}_{+,+1}(x,\textbf{k}_{\perp})\mid^2 + \mid\Psi^{\uparrow}_{+,-1}(x,\textbf{k}_{\perp})\mid^2 - \mid\Psi^{\uparrow}_{-,+1}(x,\textbf{k}_{\perp})\mid^2],\nonumber\\
g_{1T}^{e}(x,\textbf{k}_{\perp}^2) &=& \frac{M}{32\pi^3 k_{\perp}^2}\sum_{\lambda}[(k_{x}+\iota k_{y}) {\Psi^{\uparrow*}_{+,\lambda}}(x,\textbf{k}_{\perp})\Psi^{\downarrow}_{+,\lambda}(x,\textbf{k}_{\perp})
+(k_{x}-\iota k_{y}){\Psi^{\downarrow*}_{+,\lambda}}(x,\textbf{k}_{\perp})\Psi^{\uparrow}_{+,\lambda}(x,\textbf{k}_{\perp})],\nonumber\\
h_{1L}^{\bot e}(x,\textbf{k}_{\perp}^2) &=& \frac{M}{16\pi^3 k_{\perp}^2}\sum_{\lambda}[(k_{x}+\iota k_{y}){\Psi^{\uparrow*}_{-,\lambda}}(x,\textbf{k}_{\perp})\Psi^{\uparrow}_{+,\lambda}(x,\textbf{k}_{\perp})+(k_{x}-\iota k_{y}){\Psi^{\uparrow*}_{+,\lambda}}(x,\textbf{k}_{\perp})\Psi^{\uparrow}_{-,\lambda}(x,\textbf{k}_{\perp})],\nonumber\\
h_{1}^{e}(x,\textbf{k}_{\perp}^2) &=& \frac{1}{16\pi^3}[{\Psi^{\uparrow*}_{+,+1}}(x,\textbf{k}_{\perp})\Psi^{\downarrow}_{-,+1}(x,\textbf{k}_{\perp})+ {\Psi^{\uparrow*}_{+,-1}}(x,\textbf{k}_{\perp})\Psi^{\downarrow}_{-,-1}(x,\textbf{k}_{\perp})],\nonumber\\
h_{1T}^{\bot e}(x,\textbf{k}_{\perp}^2) &=& \frac{m}{16\pi^3 k_{\perp}^2}\sum_{\lambda}[(k_{x}+\iota k_{y})^2{\Psi^{\uparrow*}_{-,\lambda}}(x,\textbf{k}_{\perp})\Psi^{\downarrow}_{+,\lambda}(x,\textbf{k}_{\perp})+(k_{x}-\iota k_{y})^2 {\Psi^{\downarrow*}_{+,\lambda}}(x,\textbf{k}_{\perp})
\Psi^{\uparrow}_{-,\lambda}(x,\textbf{k}_{\perp})].\nonumber
\\
\end{eqnarray}
The expressions of T-odd TMDs $ f_{1T}^e $ and $h_{1}^{\bot e}$ vanished due to the absence of gauge link
\cite{Bacchetta}. 
\section{Simulated Model Calculations}
In this section, we have obtained the TMDs in a simulated QED model of lepton. The wave functions for the two-particle Fock state of an electron with up helicity ($ \Lambda = \uparrow$) are \cite{brodsky_a}:
\begin{eqnarray} 
\Psi^{\uparrow}_{+,+1}(x,\textbf{k}_{\perp})&=& - \sqrt{2}\frac{- k_{x} +\iota k_{y}}{x(1-x)}\varphi,  \qquad 
\Psi^{\uparrow}_{+,-1}(x,\textbf{k}_{\perp})= - \sqrt{2}\frac{k_{x} + \iota k_{y}}{(1-x)}\varphi,\nonumber\\
\Psi^{\uparrow}_{+,+1}(x,\textbf{k}_{\perp})&=& - \sqrt{2}\Big(M - \frac{m}{x}\Big)\varphi,  \qquad 
\Psi^{\uparrow}_{+,-1}(x,\textbf{k}_{\perp})= 0, 
\end{eqnarray}
  where,
\begin{eqnarray}
\varphi(x,\textbf{k}_{\perp}) = \frac{\emph{e}}{\sqrt{1-x}}\frac{1}{M^2-\frac{\textbf{k}_{\perp}^{2} + m^{2}}{x} - \frac{\textbf{k}_{\bot}^{2} + \lambda^{2}}{1-x}}.
\end{eqnarray}
Similarly for down helicity ($ \Lambda = \downarrow$), we have
\begin{eqnarray}
\Psi^{\downarrow}_{+,+1}(x,\textbf{k}_{\perp})= 0 , \qquad \qquad \qquad
\Psi^{\downarrow}_{+,-1}(x,\textbf{k}_{\perp})&=& - \sqrt{2}\Big(M - \frac{m}{x}\Big)\varphi,\nonumber\\
\Psi^{\downarrow}_{+,+1}(x,\textbf{k}_{\perp})= - \sqrt{2}\frac{-k_{x}+ \iota k_{y}}{(1-x)}\varphi, \qquad
\Psi^{\downarrow}_{+,-1}(x,\textbf{k}_{\perp})&=& - \sqrt{2}\frac{k_{x}+ \iota k_{y}}{x(1-x)}\varphi .
\end{eqnarray}
We have used the generalized form of QED by assigning a mass $M$ to the external electrons and a different mass $m$ to the internal electron and a mass $ \lambda $ to the internal photon. To derive the simulated QED model, light-front wave function $\varphi(x,\textbf{k}_{\perp})$ in Eq. (5) is differentiated  with respect to $ M^{2}$ as in \cite{asmita}. In other words, we take
\begin{eqnarray}
\varphi^{'}(x,\textbf{k}_{\perp})= M^2 \frac{\partial\varphi(x,\textbf{k}_{\perp})}{\partial M^2} = \frac{\emph{e}}{\sqrt{1-x}}\frac{M^2}{\Big(M^{2} - \frac{\textbf{k}_{\bot}^{2} + m^{2}}{x} - \frac{\textbf{k}_{\bot}^{2} + \lambda^{2}}{1-x}\Big)^2}.
\end{eqnarray}
 This improves the convergence of the wave functions at the end points of $x$ as well as the $\textbf{k}_{\perp}^{2}$ behaviour as shown in \cite{narinder}.\\
\begin{figure}
\centering
\includegraphics[width=2.5 in]{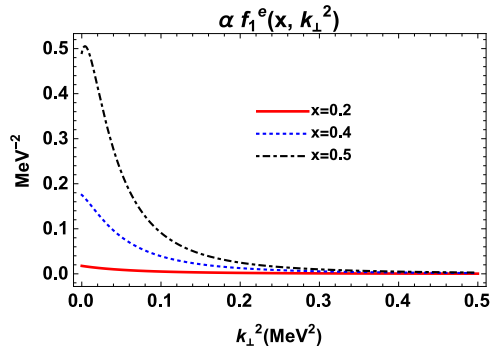}
\caption{Plot of unpolarized distribution function $f_{1}^{e}(x,k_{\perp}^2)$ as a function of $ k_{\perp}^2 $ for different values of $ x $. All distributions are rescaled by coupling constant ($ \alpha $).}
\label{fig:1}
\end{figure}
\begin{figure}
\begin{minipage}[c]{0.9\textwidth}
\includegraphics[width=2.5 in]{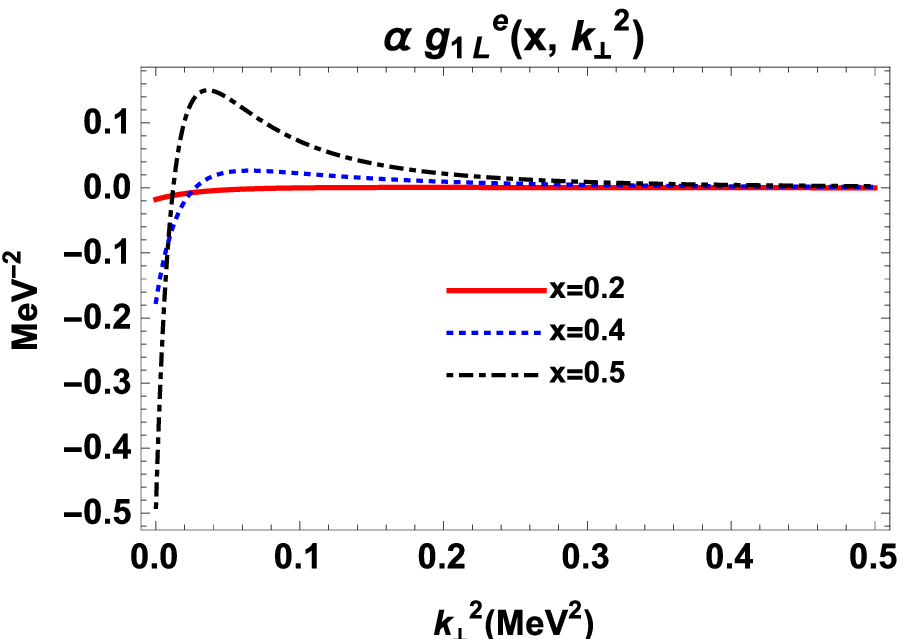}\hfill
\includegraphics[width=2.5 in]{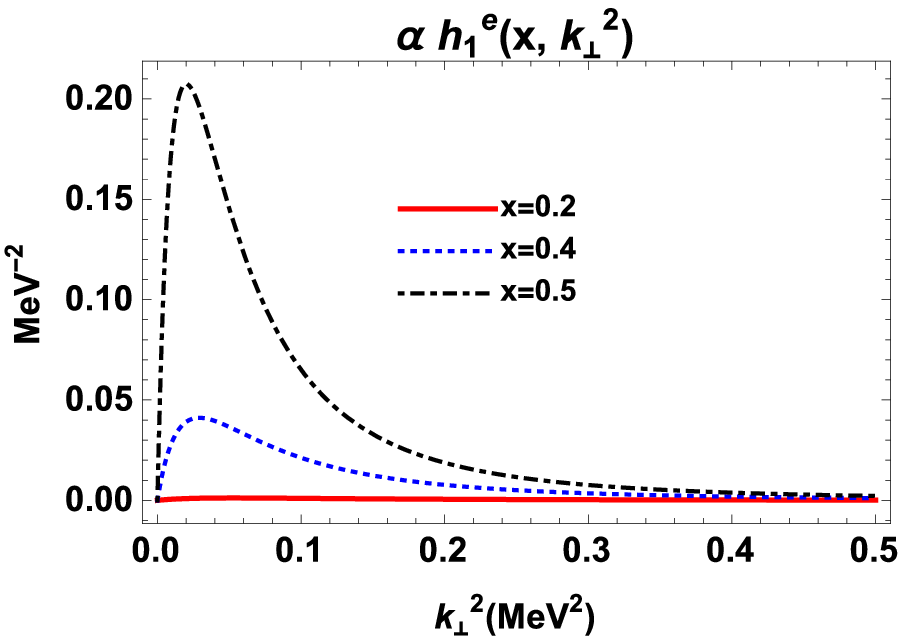}
\caption{Plots of TMDs $ g_{1L}^{e}(x,k_{\perp}^2)$ and $ g_{1T}^{e}(x,k_{\perp}^2)$, rescaled by coupling constant ($ \alpha $), as a function of $ k_{\perp}^2 $ for different values of $ x $.}
\label{fig:2}
\end{minipage}
\end{figure}
\begin{figure} 
\begin{minipage}[c]{0.9\textwidth}
\includegraphics[width=2.595 in]{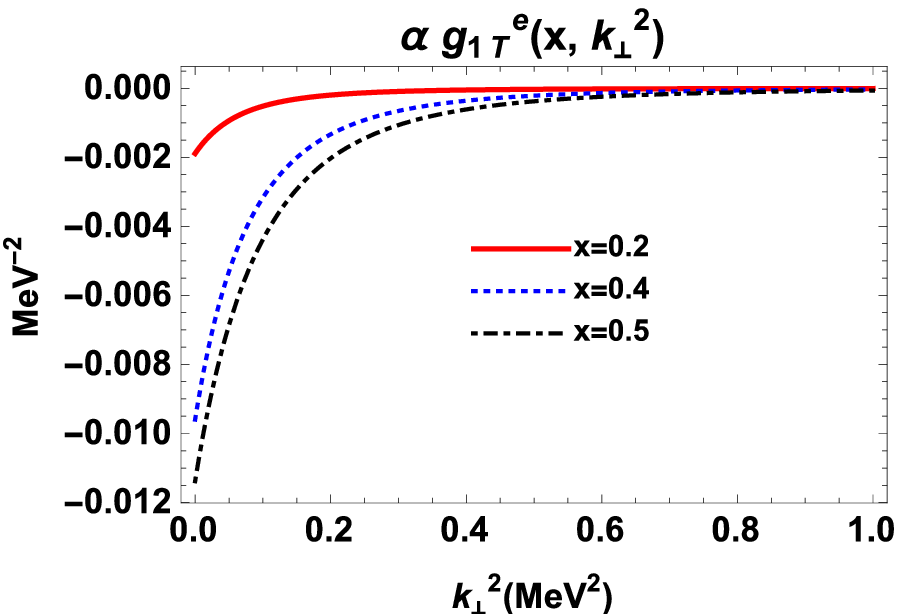}\hfill
\includegraphics[width=2.5 in]{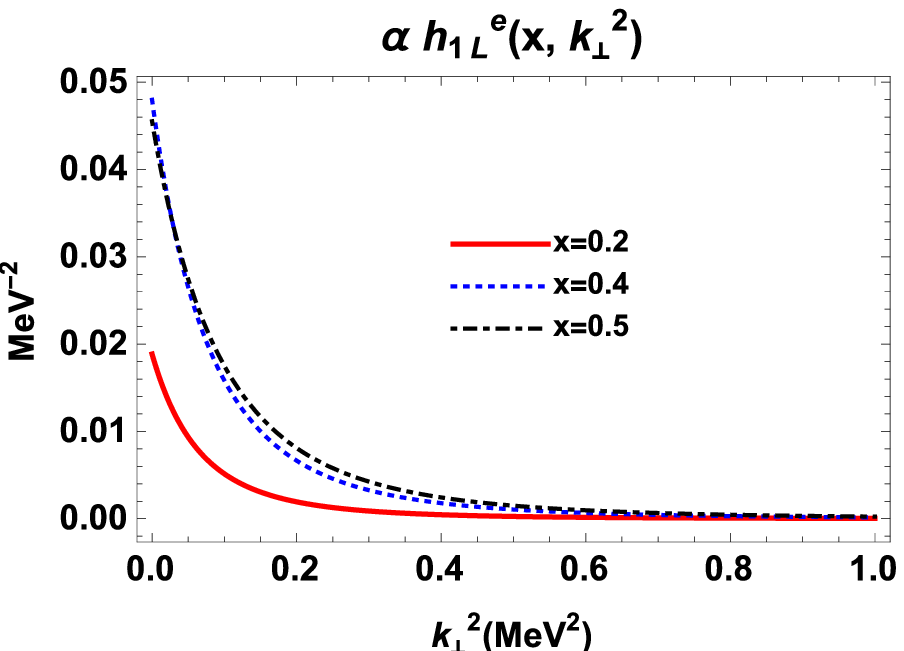}
\caption{The dependence of TMDs $h_{1L}^{e}(x,k_{\perp}^2)$ and $h_{1T}^{e}(x,k_{\perp}^2)$, rescaled by coupling constant ($ \alpha $), as a function of $ k_{\perp}^2 $ for different values of $ x $.} 
\label{fig:3}
\end{minipage}      
\end{figure}
The analytic results of T-even TMDs in this model are
\begin{eqnarray}
f_{1}^{e}(x,\textbf{k}_{\perp}^{2}) &=& \frac{e^{2}}{(2\pi)^{3}}\frac{x^{2}(1-x)M^4[k_{\perp}^{2}(1+x^2) + (1-x)^{2}(M x-m)^{2}]}{\big[ M^{2}x(1-x)-(k_{\perp}^{2}+ m^{2})(1-x)-(k_{\perp}^{2}+ \lambda^{2})x \big]^{4}} ,\nonumber\\
g_{1L}^{e}(x,\textbf{k}_{\perp}^{2}) &=& \frac{e^{2}}{(2\pi)^{3}}\frac{x^{2}(1-x)M^4[k_{\perp}^{2}(1+x^2) - (1-x)^{2}(M x-m)^{2}]}{\big[M^{2}x(1-x)-(k_{\perp}^{2}+ m^{2})(1-x)-(k_{\perp}^{2}+ \lambda^{2})x \big]^{4}},\nonumber\\
g_{1T}^{e}(x,\textbf{k}_{\perp}^{2}) &=& \frac{e^{2}}{(2\pi)^{3}}\frac{x^{3}(1-x)^{2}M^5(M x-m)}{\big[M^{2}x(1-x)-(k_{\perp}^{2}+ m^{2})(1-x)-(k_{\perp}^{2}+ \lambda^{2})x \big]^{4}},\nonumber\\
h_{1L}^{\bot e}(x,\textbf{k}_{\perp}^{2}) &=& \frac{- 2 e^{2}}{(2\pi)^{3}}\frac{x^{2}(1-x)^{2}M^5(M x-m)}{\big[M^{2}x(1-x)-(k_{\perp}^{2}+ m^{2})(1-x)-(k_{\perp}^{2}+ \lambda^{2})x\big]^{4}}.\nonumber\\
h_{1}^{e}(x,\textbf{k}_{\perp}^2) &=& \frac{e^2}{(2\pi)^3}\frac{2 x^{3}(1-x)M^4 k_{\perp}^2}{\big[M^{2}x(1-x)-(k_{\perp}^{2}+ m^{2})(1-x)-(k_{\perp}^{2}+ \lambda^{2})x \big]^{4}},\nonumber\\
h_{1T}^{\bot e}(x,\textbf{k}_{\perp}^2) &=& 0.
\end{eqnarray}
\section{Results and Discussion}
We have simulated the QED model for a physical electron 
and then obtained the results for T-even TMDs. The pzetzelosity TMD for an electron in this model also vanishes. In Fig. (1), (2)and (3), we have shown the plots of the remaining T-even TMDs as the function of $ k_{\perp}^{2} $ for a few value of $ x $.  It can be seen from the plots that at the higher values of $ k_{\perp}^{2} $, the TMDs 
become independent of $ x $ but at lower value of $ k_\perp^2$, they highly depend on the value of $ x $. 

\end{document}